\documentclass[preprint,12pt]{elsarticle}
\usepackage{graphicx}
\usepackage{amsmath}
\usepackage{amssymb}
\usepackage[nolists]{endfloat}

\journal{Journal of Computational Physics}

\begin{document}

\begin{frontmatter}

\title{Initializing and stabilizing variational multistep algorithms for modeling dynamical systems}
\author[pppl]{C. L. Ellison}
\ead{lellison@pppl.gov}
\author[pppl]{J. W. Burby}
\author[lanl]{J. M. Finn}
\author[pppl,ustc]{H. Qin}
\author[pppl]{W. M. Tang}

\address[pppl]{Princeton Plasma Physics Laboratory, Princeton, NJ 08543, USA}
\address[lanl]{Los Alamos National Laboratory, Los Alamos, NM 87545, USA}
\address[ustc]{Department of Modern Physics, University of Science and Technology of China, Hefei, Anhui 230026, China}

\begin{abstract}
Backward error initialization and parasitic mode control are well-suited for use in algorithms that arise from a discrete variational principle on phase-space dynamics. Dynamical systems described by degenerate Lagrangians, such as those occurring in phase-space action principles, lead to variational multistep algorithms for the integration of first-order differential equations. As multistep algorithms, an initialization procedure must be chosen and the stability of parasitic modes assessed. The conventional selection of initial conditions using accurate one-step methods does not yield the best numerical performance for smoothness and stability. Instead, backward error initialization identifies a set of initial conditions that minimize the amplitude of undesirable parasitic modes. This issue is especially important in the context of structure-preserving multistep algorithms where numerical damping of the parasitic modes would violate the conservation properties. In the presence of growing parasitic modes, the algorithm may also be periodically re-initialized to prevent the undesired mode from reaching large amplitude.  Numerical examples of variational multistep algorithms are presented in which the backward error initialized trajectories outperform those initialized using highly accurate approximations of the true solution. 
\end{abstract}

\begin{keyword}
Variational Integrators \sep Multistep Algorithms \sep Backward Error Analysis
\end{keyword}

\end{frontmatter}

\section{Introduction}

The primary advantage of multistep algorithms is the ability to achieve accuracy greater than first-order in step size $h$ while requiring only a single additional function evaluation at each time step. Multistep methods are thus a common technique for the numerical solution of ordinary differential equations (ODEs), encompassing the well-known Adams-Bashforth and Adams-Moulton families of algorithms \cite{Dahlquist_1956, Hairer_1993, Hairer_2006}. The presence of additional steps in the update rule, however, leads to unique considerations for stability and initialization \cite{Hairer_2006, Hairer_1999}. Generating the additionally-required initial conditions using an arbitrary one-step method may not achieve the desired accuracy \cite{Tirani_2003} or yield sufficiently smooth solutions \cite{Hairer_2006, Hairer_1999, DAmbrosio_2013}. Multistep algorithms also exhibit parasitic modes -  unphysical behavior originating from eigenvalues that do not lie near $1$ on the unit circle \cite{Hairer_2006, Hairer_1999, DAmbrosio_2013, Hairer_2004, Hairer_2006_slmm, Console_2012, Console_2013}. Often these parasitic modes are apparent when computing conserved quantities, manifesting as even-odd or n-cycle oscillations imposed upon some smoother trend. Left unmitigated, the parasitic mode amplitude can grow to dominate the numerical error. The potential pitfalls of parasitic modes may suggest only strictly stable multistep algorithms should be used to ensure all of the parasitic modes decay with time. However, important structure-preserving classes of algorithms exhibit marginally stable parasitic modes \cite{Hairer_2006}, and the numerical advantages of conservative algorithms may outweigh the disadvantages of parasitic mode mitigation.  

One context in which conservative multistep methods naturally emerge is variational integration \cite{Marsden_2001} of dynamical systems described by degenerate Lagrangians \cite{Rowley_2002, Vankerschaver_2013, Qin_2008, Qin_2009, Li_2011, Squire_2012, Ober-Blobaum_2013, Yoshimura_2006_I, Yoshimura_2006_II, Burby_2014}. Degenerate Lagrangians appear in contexts in which constraints are present, such as electric circuit models \cite{Yoshimura_2006_E}, and in differential equations resulting from Hamilton's principle on phase space \cite{Arnold_1989}. These phase-space action principles appear in the descriptions of interacting point vortices \cite{Aref_2007}, the guiding-center motion of charged particles in magnetic fields \cite{Littlejohn_1983, Cary_2009}, and Hamiltonian descriptions of magnetic field line flow \cite{Cary_1983}. Unless the phase-space action principle can be represented in canonical coordinates \cite{Leok_2011, Lall_2006}, it is unknown how to discretize Hamilton's principle on phase space in such a manner as to obtain one-step numerical methods. Harnessing the benefits of structure-preserving variational algorithms in systems with degenerate Lagrangians therefore requires consideration of the multistep aspect of the algorithms.

Backward error analysis is a powerful tool for understanding the behavior of multistep methods and identifying initial conditions which lead to smooth numerical behavior. The backward error analysis for multistep algorithms is presented by Hairer in Ref.\,\cite{Hairer_1999}, with similar discussions present in the textbook by Hairer, Lubich, and Wanner \cite{Hairer_2006}. By seeking a ``modified" differential equation which is nearby the physical system being modeled, comparison of the modified system to the original illuminates features that are present in the numerical dynamics but not the physical flow. For multistep methods, a ``smooth" modified system emerges by assuming the absence of parasitic modes, thereby describing the time evolution of the desired component of the numerical solution. To explain the more general behavior of the multistep method, it is necessary to consider a ``full" modified system that also accounts for the parasitic modes. The nonlinear parasitic mode stability and initial parasitic mode amplitude may be deduced from the full modified system. The minimum parasitic mode amplitudes occur when initial conditions lie along solutions to the smooth modified system, and sampling initial conditions along solutions to truncations of the smooth modified system can greatly improve the resulting numerical behavior \cite{DAmbrosio_2013}.

In this paper, we illustrate the benefits of initializing multistep variational integrators using initial conditions sampled along solutions of the smooth modified system described by Hairer \cite{Hairer_1999}. This ``backward error initialization" procedure retains several terms in the smooth modified system to reduce the initial parasitic mode amplitude. Numerical examples demonstrate the advantages compared to standard generation of initial conditions for multistep algorithms. To further mitigate the growing unphysical modes that can arise in the numerical trajectories, we make the novel suggestion of performing parasitic mode control by periodically re-initializing the algorithm using the backward error procedure. In the past, parasitic mode control has been performed by advancing partial steps with Runge-Kutta algorithms and averaging away the predominantly undesirable behavior \cite{Aoyagi_1991, New_1998}. Compared with Runge-Kutta based smoothing, backward error re-initialization yields points that nearly lie on or near a trajectory of the smooth modified system. While the re-initialization step is not structure preserving, the parasitic modes are restarted with minimal amplitude, allowing many steps of the standard algorithm between re-initialization steps. Finally, formal significance of the smooth modified system flow is established by identifying a symplectic structure which is preserved on the original phase space instead of the standard structure which exists in a space of larger dimension.

The organization of the manuscript is as follows. In Section \ref{sec:initialization_method}, the backward error initialization method is reviewed for simple multistep algorithms and its relevance to variational multistep methods established. Section \ref{sec:parasitic_mode_control} discusses the control of parasitic modes by monitoring and re-initializing the undesired modes.  In Section \ref{sec:numerical_examples}, numerical examples are presented that demonstrate the utility of backward error initialization and parasitic mode control. Each of the example algorithms emerge from a discrete variational principle, but are intentionally chosen to be familiar from a more conventional multistep perspective. Section \ref{sec:conclusion} delivers closing remarks.

\section{Initialization using Backward Error Analysis}
\label{sec:initialization_method}

In this section, we describe the selection of initial conditions dictated by backward error analysis and motivated by reducing parasitic mode amplitudes observed in variational multistep algorithms. The backward error analysis follows the work of Ref.\,\cite{Hairer_1999}. First the initial condition selection procedure is presented, then discussions provide context for the smooth modified system of multistep algorithms. The ideal multistep trajectory given by the flow of the smooth modified system is shown to conserve a symplectic structure on the original phase space. This is a stronger result than the standard symplectic structure preserved by multistep variational integrators \cite{Marsden_2001, Rowley_2002, Squire_2012}, which is on a product of the phase space with itself.

\subsection{Selecting Initial Conditions}
\label{subsec:selecting_initial_conditions}

Consider an initial value problem specified by a first-order differential equation in the form:
\begin{equation}
	\dot{x} = f(x),
	\label{eq:original_ode}
\end{equation}
with initial condition $x(t=0) = x_0$. In our cases of interest, this ODE will be in Hamiltonian form and is assumed time-independent for simplicity. We seek to model the initial value problem using a multistep numerical method with time increments of size $h$. That is, we approximate the continuous solution $x(t)$ at discrete times $t=0, h, ..., Nh$, and the notation $x_j$ is used to denote the numeral approximation of $x(t=jh)$. The $k$-step multistep method is then an $h$-dependent map $\Phi_h(x_{n}, ..., x_{n+k-1}) = x_{n+k}$. Let the \textit{update rule} be the real-valued function $F_h(x_{n}, ..., x_{n+k})$ such that $F_h(x_{n}, ..., \Phi_h(x_{n}, ..., x_{n+k-1})) = 0$. That is, $F_h$ is zero when a set of $k$ values satisfies the numerical algorithm. For the backward error analysis that follows, we will restrict ourselves to linear multistep methods where the update rule takes the form:
\begin{equation}
	F_h(x_{n}, ..., x_{n+k}) = \sum_{j=0}^k \alpha_j x_{n+j} - h \beta_j f(x_{n+j}) = 0,
	\label{eq:linear_multistep_method}
\end{equation}
where $\alpha_j, \beta_j$ are real parameters such that $\alpha_k \ne 0$, $\left|\alpha_0\right| + \left|\beta_0 \right| > 0$. While not all variational multistep algorithms are linear in this sense, two of the examples in Sec.\,\ref{sec:numerical_examples} are of this form and the simplicity of notation over treatment of more general multistep algorithms is advantageous for familiarizing the reader with multistep backward error analysis and initialization.

Following Refs.\,\cite{Hairer_2006, Hairer_1999, Hairer_2004}, the \textit{smooth modified system} is the unique ODE of the form:
\begin{equation}
  \dot{\tilde{x}} = f(\tilde{x}) + h f_2(\tilde{x}) + h^2 f_3(\tilde{x}) + ... ,
	\label{eq:smooth_modified_system}
\end{equation}
such that, for any truncation order $N$,  any solution $\tilde{x}(t)$ of the smooth modified system retaining terms up to $f_N$ satisfies $F_h(\tilde{x}((n-k)h), ..., \tilde{x}(nh)) = \mathcal{O}(h^{N+1})$. In the preceding definition, $f(\tilde{x})$ must be that of Eq.\,(\ref{eq:original_ode}) for consistency and the $f_i(\tilde{x})$ are smooth, h-independent functions. The coefficients $f_i$ must be recursively determined by the update rule and original ODE. The modified system is a possibly non-convergent asymptotic series, requiring truncation after an appropriate number of terms to obtain minimum error \cite{Hairer_2006}. For the purposes of generating practical initial conditions, the optimal truncation order is likely to be greater than the chosen truncation order, so we defer convergence discussion to the References \cite{Hairer_2006,Hairer_1999}. Note that because Eq.\,(\ref{eq:smooth_modified_system}) is a first order ODE of the same dimension as that of the original system, solutions to the smooth modified system are uniquely identified by a single initial condition.

\textit{Backward error initialization} is given by the following procedure:
\begin{enumerate}
 \item{Assume a smooth modified system of the form Eq.\,(\ref{eq:smooth_modified_system}).}
 \item{Taylor expand the update rule $F_h$ about $t_0$, substituting the smooth modified system for the time derivatives of $x$.}
 \item{Determine the unknown functions $f_i$ by collecting in orders of $h$ in the expanded update rule.}
 \item{Retain one or more non-zero coefficients $f_i$ to identify a truncated smooth modified system.}
 \item{Use an accurate one-step method to solve the initial value problem defined by the truncated smooth modified system and the single initial condition $x_0$. Advance forward in time for $k-1$ steps.}
 \item{Supply the resulting numerical trajectory $\tilde{x}_0, \tilde{x}_1, ..., \tilde{x}_{k-1}$ as the set of initial conditions for the $k$-step method.} 
\end{enumerate}
The critical distinction between backward error initialization and conventional multistep initialization is to seek the best approximation of the solution to the smooth modified system rather than the original ODE.

At this point, it is important to note that while solutions of the (truncated) smooth modified system Eq.\,(\ref{eq:smooth_modified_system}) satisfy the update rule (to one order in $h$ greater than the truncation order), arbitrary numerical trajectories obtained through iteration of the numerical map do not necessarily lie on or near solutions of the smooth modified system. In essence, the ansatz that the numerical solution behaves like the smooth function $\tilde{x}(t)$ is not sufficiently general to capture the full behavior realizable by the numerical algorithm. Because initial conditions do not typically lie along solutions to the smooth modified system, a more general modified system is required to account for the observed numerical behavior.

\subsection{Full Modified System}

Instead, one must make the more general ansatz that the numerical trajectory possesses some number of non-smooth modes. The ODE describing the time evolution of the functions appearing in the more general ansatz is denoted the ``full" modified system in contrast to the smooth modified system. To derive the full modified system, suppose the solution takes the form \cite{Hairer_2006}:
\begin{equation}
	\tilde{x}(t) = \tilde{y}(t) + \sum_{\zeta_l \in \mathcal{I}} \zeta_l^{t/h} \tilde{z}_l(t), 
        % Is this ok? Can you sum over \zeta_l and imply the l in z_l is the same?
	\label{eq:full_modified_solution}
\end{equation}
where $\tilde{y}(t), \tilde{z_l}(t)$ are smooth functions and $\mathcal{I}$ is a set of points on the unit circle that depends on the multistep method. The set $\mathcal{I}$ is typically determined by solving for the eigenvalues of the algorithm applied to the trivial function $f(x) = 0$ and forming all possible combinations of those eigenvalues. That is, let $h\rightarrow 0$ in the update rule and solve the characteristic polynomial equation $F_{h=0}( 1, \zeta, ..., \zeta^{n-1}) = 0$. A multistep method is deemed ``zero-stable" if $\left|\zeta\right| \leq 1$ for all of these roots. Symmetric multistep methods are zero-stable if and only if $\left| \zeta \right| = 1$ for all $\zeta$. Assuming the algorithm is consistent, multistep methods will possess a root at $\zeta=1$, denoted the principle root, and $k-1$ distinct ``parasitic" roots $\zeta_{l=2, ..., k}$ that lie on or within the unit circle away from $1$. The set $\mathcal{I}$ is composed of all $N$-multicombinations of the parasitic roots with finite $N$ and excluding the values at $1$. In certain circumstances, one must include finite $h$ terms in the characteristic polynomial \cite{Hairer_Lubich_2013}. For variational multistep methods, assuming the algorithm is linearly stable, all parasitic roots will lie on the unit circle. For the two-step methods of interest in the numerical examples, $\mathcal{I} = \{-1\}$. Also note that our definition of $\mathcal{I}$ is slightly different from that of Ref.\,\cite{Hairer_2006}. 

The full modified system describing the evolution of the ansatz Eq.\,(\ref{eq:full_modified_solution}) is given by Theorem XV.3.5 in \cite{Hairer_2006}, which says that for any truncation order $N$, any solution of:
\begin{align}
	\dot{\tilde{y}} & = f_{1,1}(\tilde{x}) + h f_{1,2}(\tilde{x}) + ... + h^{N-1} f_{1, N-1}(\tilde{x}) \nonumber \\
	\dot{\tilde{z_l}}& = f_{l,1}(\tilde{x}) + h f_{l,2}(\tilde{x}) + ... + h^{N-1} f_{l, N-1}(\tilde{x}) \quad \text{for} \quad 2 \le l \le k \nonumber \\
	\tilde{z_l} & = h f_{l,2}(\tilde{x}) + ... + h^N f_{l, N+1}(\tilde{x}) \quad \text{for} \quad l > k,
	\label{eq:full_modified_system}
\end{align}
satisfies the update rule to order $\mathcal{O}(h^{N+1})$, assuming the initial conditions are consistent with the original ODE. 

The solution ansatz Eq.\,(\ref{eq:full_modified_solution}) and full modified system Eq.\,(\ref{eq:full_modified_system}) clarify the presence of parasitic modes and the impact of the excess initial conditions on multistep method trajectories. The equation describing $\dot{\tilde{y}}$ in Eq.\,(\ref{eq:full_modified_system}) is strongly related to the smooth modified system Eq.\,(\ref{eq:smooth_modified_system}), however, different $f_i$ coefficients can appear when $\zeta_i^m \zeta^n_j = 1$ for some powers $m,n$. In contrast to the smooth modified system, the full modified system is of dimension $k$-times larger than the original system. The number of initial conditions required to specify a solution to the full modified system matches the number of initial conditions required to begin using the multistep method. To the extent which a set of initial conditions does not lie on a solution to the smooth modified system, the parasitic modes $z_l$ will be initialized with non-zero amplitude. In particular, initial conditions that exactly satisfy a smooth modified system truncated at order $\mathcal{O}(h^{N-1})$ will yield parasitic modes of initial amplitude scaling as $\mathcal{O}(h^{N+1})$. Initial conditions sampled from the true solution yield parasitic modes of initial amplitude scaling as $\mathcal{O}(h^{m+1})$, where $m$ is the accuracy order of the multistep method, due to the agreement of the smooth modified system up to the first non-zero $f_i$ at $i=(m+1)$. 

\subsection{Smooth Modified System for Linear Multistep Methods}

As an example calculation relevant to several of the numerical examples, suppose the multistep method is linear as in Eq.\,\ref{eq:linear_multistep_method}. Calculation of the smooth modified system then follows the discussion of \cite{Hairer_1999} and is reproduced here for the convenience of the reader. Equipping ourselves with the smooth modified system Eq.\,(\ref{eq:smooth_modified_system}), we assume $f$ is sufficiently continuous that the derivatives exist and are smooth at the needed orders. The goal is to recursively solve for the unknown functions $f_i(x)$ for $i \geq 2$. We will do so by Taylor expanding the terms appearing in the multistep method, substituting the modified system for the time derivatives of $x$, and matching by orders of $h$. It is then helpful to expand $\tilde{x}(t_0 + jh)$ for some integer $j$:
\begin{align}
  \tilde{x}(t_0 + jh) & = \tilde{x}_0 + j h \dot{\tilde{x}}_0 + \frac{j^2h^2}{2} \ddot{\tilde{x}}_0 + ... \nonumber \\
  & = \tilde{x}_0 + jh (f(\tilde{x}_0) + h f_2(\tilde{x}_0) + h^2 f_3(\tilde{x}_0) + ... ) + \nonumber \\
  & \quad \quad \quad \  \frac{j^2h^2}{2} (f'(\tilde{x}_0) + h f_2'(\tilde{x}_0) + ...)(f(\tilde{x}_0) + h f_2(\tilde{x}_0) + ... )
  \label{eq:expanded_x_tilde}
\end{align}
where the zero subscript denotes evaluation at time $t=0$ and we have substituted the smooth modified system for the time derivatives. We similarly expand the $f$ evaluations that will appear in the multistep method:
\begin{align}
  f(\tilde{x}(t_0 + jh)) & = f(\tilde{x}_0) +  f'(\tilde{x}_0)(\tilde{x}(t_0 + jh) - \tilde{x}_0) + \nonumber \\
 & \quad \quad \ \ f''(\tilde{x}_0)(\tilde{x}(t_0 + jh) - \tilde{x}_0, \tilde{x}(t_0 + jh) - \tilde{x}_0) + ... \nonumber \\
 	& = f + f'[jh(f +  h f_2 + ...) + \frac{j^2 h^2}{2} (f' + ...)(f + ...) + ...] \nonumber \\
 		& \quad \quad  + f''[jh(f + ...) + ...][jh(f + ...) + ...] + ... ,
  \label{eq:expanded_f_of_x_tilde}
\end{align}
where all functions evaluations occur at $\tilde{x}_0$ in the final line.

Substituting Eq.\,(\ref{eq:expanded_x_tilde}) for the $\alpha$ terms and Eq.\,(\ref{eq:expanded_f_of_x_tilde}) for the $\beta$ terms in the linear multistep method Eq.\,(\ref{eq:linear_multistep_method}), we collect in orders of $h$. The first two powers of $h$ yield consistency relations:
\begin{align}
	h^0 & :  \quad \quad \quad \quad \sum_{j=0}^k \alpha_j = 0 \nonumber \\
	h^1 & : \quad \quad \sum_{j=0}^k j \alpha_j - \beta_j = 0. 
\end{align}
With the added assumption that the algorithm is normalized, namely $\sum_j j \alpha_j = \sum_j \beta_j = 1$, the $h^2$ term defines $f_2$:
\begin{equation}
	f_2(\tilde{x}) = \frac{-1}{2} \sum_{j=0}^k (j^2 \alpha_j - 2 j \beta_j) f'(\tilde{x}) f(\tilde{x}).
\end{equation}
For symmetric algorithms $f_2$ and indeed $f_i$ for any even $i$ is zero. The first potentially non-zero term of the modified system for symmetric methods occurs at order $\mathcal{O}(h^2)$: 
\begin{align}
	f_3(\tilde{x}) = & \frac{1}{3!} \left\{  f''(\tilde{x}) f(\tilde{x}) f(\tilde{x}) \left[(\sum_j 3j^2\alpha_j)(\sum_j\frac{j^2 \alpha_j}{2} - j\beta_j) + \sum_j 3j^2 \beta_j - j^3 \alpha_j \right] +  \right. \nonumber \\
	 &  \left.  f'(\tilde{x}) f'(\tilde{x}) f(\tilde{x}) \left[ 6(\sum_j j^2\alpha_j - j \beta_j)(\sum_j\frac{j^2 \alpha_j}{2} - j\beta_j) + \sum_j 3 j^2 \beta_j - j^3 \alpha_j \right] \right\}. 
\end{align}
One may recursively solve for increasing orders in the modified system in terms of the original function $f$, its derivatives, and the constants defining the multistep method. Compact representations of these expressions are derived in Ref.\,\cite{Hairer_1999} using B-series and rooted trees, and an example routine is presented in Chapter IX of Ref.\,\cite{Hairer_2006} for calculating these functions using a symbolic package. We reproduce the explicit definitions for convenience and clarity in the numerical examples. 

Recalling that solutions of the smooth modified system truncated at order $\mathcal{O}(h^{N-1})$ result in parasitic modes with initial amplitudes scaling like $\mathcal{O}(h^{N+1})$, a significant impact on the numerical trajectory results from solving the truncated smooth modified system forward in time to generate the initial conditions. Numerical examples will be shown in Section \ref{sec:numerical_examples}. For symmetric linear multistep methods, because all odd $f_i$ are zero, retaining the correction $f_3$ yields parasitic modes that scale as $\mathcal{O}(h^5)$. In contrast, initialization with the true solution exhibits $\mathcal{O}(h^3)$ parasitic modes. To implement these improved initial conditions, one must provide the Hessian and Jacobian of the ODE function $f$ and an implementation of an accurate one-step method such as fourth-order Runge-Kutta or implicit midpoint. The Jacobian is likely to already have been implemented for the nonlinear solution of implicit multistep methods, and use of symbolic packages or automatic differentiation tools can calculate the second order derivatives with minimal effort.

Given the end goal of using the \emph{solution} of the smooth modified system, one may naively attempt a forward error analysis to bypass the smooth modified system calculation. Namely, one could attempt an expansion of the solution $\tilde{x}(t)$ in powers of $h$ and recursively solving for the unknown functions. This approach dates back to the so-called ``underlying one-step method" of Kirchgraber \cite{Kirchgraber_1986}, known in other literature as the ``step transition operator" \cite{Feng_1998}. In essence, seeking an operator that maps from the single point $x_j$ to the next point $x_{j+1}$, and after iterating $k$ times the set of points is a solution to the $k$-step method. While it is, essentially, the underlying one-step method that we are attempting to approximate by determining the smooth modified system, the modified system typically exhibits better convergence properties when ordered by step size $h$ than the underlying one-step method \cite{Hairer_2006}. Thus it is more robust to truncate and numerically solve the smooth modified system than it is to directly seek the underlying one-step method.

\subsection{Smooth Modified System for Phase Space Variational Integrators}
\label{ssec:smooth_modified_system_for_phase_space_variational_integrators}

In the context of variational integrators applied to phase-space action principles, the smooth modified system flow preserves a symplectic structure on the original phase space. That is, while the multistep variational integrator in more generally preserves a two-form on a higher dimensional space, the idealized trajectory satisfying the smooth modified system does preserve a quantity on the  phase space of interest, as will be shown in the remainder of this section.

For comparison, it is important to review the conservation properties of the continuous system whose dynamics are governed by the phase-space action principle. Suppose a  phase space $Q$ of dimension $2n$. Let local coordinates be denoted $z_1, ..., z_{2n}$. The phase-space Lagrangian is a map $L:TQ \rightarrow \mathbf{R}$, given in coordinates by:
\begin{equation}
  L(z, \dot{z}) = \vartheta_i \dot{z}^i - H(z),
\end{equation}
where $\vartheta = \vartheta_i dz^i$ is a one-form on $Q$ and the Lagrangian has been assumed to not depend explicitly on time for simplicity. The action functional $S$ is given by:
\begin{equation}
  S(z(t),\dot{z}(t)) = \int L(z(t), \dot{z}(t)) dt.
\end{equation}
Stationarity of the action integral yields (possibly non-canonical) Hamilton's Equations:
\begin{equation}
  \Omega \dot{z} = \frac{\partial H}{\partial z},
  \label{eq:noncanonical_hamiltons_equations}
\end{equation}
where $\Omega = -d \vartheta$. Solutions of the non-canonical Hamilton's equations preserve $\Omega$. That is, letting $\varphi_\tau$ be the time $\tau$ flow map associated to the Hamiltonian vector field in Eq.\,\ref{eq:noncanonical_hamiltons_equations}, $\varphi^*\Omega = \Omega$. When the phase-space Lagrangian takes the canonical form $z = [p, q]^T, \vartheta = [0, p]$, $\Omega$ reduces to the canonical symplectic structure $dp \wedge dq$. In more general coordinates, $\Omega = \vartheta_{j,i} dz^i \wedge dz^j$. Importantly, notice $\Omega$ is a two-form on the phase space $Q$. 

In constructing a variational integrator by discretizing the phase-space action principle, one might originally aspire to preserve $\Omega$ at discrete times $t_k$. However, the standard discretization procedure preserves a symplectic structure on a space \emph{twice} the dimension of the original configuration space. In particular, if $L_d(z_k, z_{k+1}) \approx \int_{kh}^{(k+1)h} L(z(t), \dot{z}(t)) dt$, the discrete flow map given by the discrete Euler-Lagrange equations preserves $\Omega_d = \frac{\partial L_d(z_k, z_{k+1})}{\partial z^i_k \partial z^j_{k+1}} dz_k^i \wedge dz_{k+1}^j$ \cite{Marsden_2001}. Previous investigations of variational integrators on phase-space Lagrangians have related this symplectic structure on $Q \times Q$ to the continuous structure on $Q$ by taking the limit of zero step size $h \rightarrow 0$ \cite{Squire_2012, Rowley_2002}. However, the notion of an idealized trajectory satisfying the smooth modified system suggests the underlying one step method of the multistep variational integrator may likely preserve a structure on $Q$. Establishing this correspondence is an important first step for obtaining rigorous results regarding the long-term behavior of multistep variational integrators. Standard proofs that symplectic algorithms exhibit bounded energy error for exponentially long times require the discrete symplectic structure to be on the original phase-space. The good long-term energy behavior observed in multistep variational integrators \cite{Rowley_2002, Qin_2008, Qin_2009, Li_2011} is then likely due to cases where parasitic modes remain small and numerical trajectories remaining close to the flow of the smooth modified system.

We assume the existence of an underlying one-step method to the multistep variational algorithms, considered to be the flow of the continuous vector field that is approximated to arbitrary accuracy in $h$ by the smooth modified system. If this map or vector field do not in fact exist, the results of this section need to be interpreted in an asymptotic sense. That is, one may only be able to preserve the symplectic structure to arbitrary order accuracy in numerical step size $h$.

To investigate the conservation properties of the solution to the smooth modified system, we restrict the discrete path space to solutions of the smooth modified system and propagate this trajectory through the action principle. That is, suppose a consistent discretization of the phase-space Lagrangian $L$ denoted by $L_d(z_k, z_{k+1})$. Also, let $\phi_\tau$ be the time $\tau$ flow map corresponding to the vector field that the asymptotic smooth modified system approximates. Starting with a single initial condition $z_0$, if $z_1= \phi_h(z_0), z_2 = \phi^2_{h}(z_0)$, then $z_0, z_1, z_2$ satisfy the discrete Euler-Lagrange equations specifying the multistep method. 

Next, the discrete action for a discrete Lagrangian is given by:
\begin{equation}
  S_d(z_0, ..., z_N) = \sum_{k=0}^{N-1} L_d(z_k, z_{k+1}).
\end{equation}
Let $\bar{S}$ be the discrete action after restricting the path space to solutions of the smooth modified system:
\begin{equation}
  \bar{S}_d(z_0) = S_d(z_0, \phi_h(z_0), ..., \phi_{Nh}(z_0)) = \sum_{k=0}^{N-1} L_d(\phi_{kh}(z_0), \phi_{(k+1)h}(z_0)),
\end{equation}
where we can identify the discrete path of $\bar{S}_d$ with the single initial condition that uniquely identifies the rest of the discrete trajectory. Applying the variational principle and letting $v_{z_0}$ be an arbitrary tangent vector at $z_0$:
\begin{align}
	d\bar{S}_d(z_0) \left[v_{z_0}\right] & = \sum_{k=0}^{N-1} d_1 L_d(\phi_{kh}(z_0), \phi_{(k+1)h} (z_0)) \left[T \phi_{kh}(v_{z_0}) \right]  +  \nonumber \\
	& \quad \quad d_2 L_d(\phi_{kh}(z_0), \phi_{(k+1)h} (z_0)) \left[T \phi_{(k+1)h} (v_{z_0})\right],
\end{align}
where $d_i$ is the exterior derivative with respect to the $i$-th variable and $T\phi : TQ \rightarrow TQ$ is the tangent map of $\phi$ \cite{Abraham_1987}. Re-arranging the summation:
\begin{align}
 d\bar{S}_d(z_0)(v_{z_0}) & = \sum_{k=1}^{N-1} \left[d_2 L_d(\phi_{(k-1) h}(z_0), \phi_{kh}(z_0)) + d_1 L_d(\phi_{kh}(z_0), \phi_{(k+1)h}(z_0)) \right] [T \phi_{kh}(v_{z_0})]  \nonumber \\
 	& \quad \quad + d_1 L_d(z_0, \phi_h(z_0)) [v_{z_0}] + d_2 L_d(\phi_{(N-1)h}(z_0), \phi_{Nh}(z_0)) [T\phi_{Nh} (v_{z_0})].
	\label{eq:intermediate_reduced_action}
\end{align}  
The summation vanishes because the discrete trajectory satisfies the discrete Euler-Lagrange Equations by construction. At this point, it is also helpful to define a pair of one-forms on $Q$:
\begin{align}
	\vartheta_a(z) [v_z] & = d_1 L_d(z, \phi_h(z))[v_z] \nonumber \\
	\vartheta_b(z) [v_z] & = d_2 L_d(\phi_{-h}(z), z)[v_z].
\end{align}
Using the one-forms $\vartheta_a, \vartheta_b$, Eq.\,(\ref{eq:intermediate_reduced_action}) reduces to:
\begin{align}
 d\bar{S}_d(z_0)(v_{z_0}) & = \vartheta_a(z_0)[v_{z_0}] + (\phi_{Nh}^*\vartheta_b)(z_0) [v_{z_0}] \nonumber \\
 & = (\vartheta_a + \phi_{Nh}^*\vartheta_b)(z_0)[v_{z_0}].
\end{align}
Because the sequence of points $\phi_{-h}(z_0), z_0, \phi_h(z_0)$ satisfies the discrete Euler-Lagrange equations, $\vartheta_b(z_0) = -\vartheta_a(z_0)$. Applying an additional exterior derivative yields:
\begin{equation}
	 \Omega_d = \phi_{(N-1)h}^* \Omega_d,
\end{equation}
where $\Omega_d = d\vartheta_a$ is given in coordinates by:
\begin{equation}
	\Omega_d = \frac{\partial^2 L_d(z, \phi_h(z))}{\partial q^i_0 \partial q_1^j} \frac{\partial \phi_h^j}{\partial z^k_0} dz^k \wedge dz^i.
\end{equation}
Here $q_0, q_1$ identify the first and second arguments of $L_d$, respectively. 
	 
The identification of a conserved quantity on the original phase space $Q$ lends additional merit to the solution of the smooth modified system as significant in the context of phase space variational integrators. While any practical implementation will deviate from the flow of the smooth modified system, the identification of a conserved quantity on $Q$ helps to relate discrete phase space variational integrators to their continuous counterpart. 

\section{Parasitic Mode Control}
\label{sec:parasitic_mode_control}

Preferably, the parasitic modes of the symmetric multistep method are oscillatory with small amplitude. Backward error initialization may then be applied once at the beginning of the simulation to yield a smooth numerical trajectory. In other circumstances, users may seek the advantages of symmetric or symplectic multistep algorithms in potentially unstable systems. The nonlinear stability of the parasitic modes requires analysis of the complicated full modified system, making it advantageous to establish a mechanism for controlling slowly growing parasitic modes. Here we suggest a method for actively monitoring and damping the parasitic mode. While the mode-reducing re-initialization step is not structure preserving, the high accuracy to which the mode is eliminated minimizes the frequency such a step must be performed.

Precedence exists for monitoring and actively smoothing parasitic mode behavior \cite{New_1998, Aoyagi_1991}. The method in Ref.\,\cite{New_1998}, for instance, monitors the parasitic behavior of the explicit midpoint algorithm until a pre-determined threshold is crossed. The offending data point is then advanced a half-step backward in time using second order Runge-Kutta while the data point preceding the threshold crossing is advanced a half-step forward:
\begin{align}
	x_{k-1/2}^+ = \Phi_{-h/2}(x_{k}) \nonumber \\
	x_{k-1/2}^- = \Phi_{h/2}(x_{k-1}).
	\label{eq:new_delousing}
\end{align}
Here, $\Phi_h$ is the numerical map generated by the Runge-Kutta algorithm with numerical step size $h$. These two approximations for $x_{k-1/2}$ are then averaged in an attempt to smooth the oscillatory even-odd behavior:
\begin{equation}
	x_{k-1/2}^s = \frac{1}{2}(x_{k-1/2}^+ + x_{k-1/2}^-).
\end{equation}
Finally, one can return to using the two-step algorithm by extrapolating the smoothed position forward and backward to the original points in time:
\begin{align}
	x_{k}^s = \Phi_{h/2}(x_{k-1/2}^s) \nonumber \\
	x_{k-1}^s = \Phi_{-h/2}(x_{k-1/2}^s),
\end{align}
where the $s$ superscript stands for ``smoothed''. 

We suggest controlling growing parasitic modes instead by re-initializing the algorithm using the backward error initialization procedure once the mode amplitude exceeds the pre-determined threshold. Compared with the Runge-Kutta stepping and averaging procedure, backward error re-initialization minimizes the parasitic mode amplitude and therefore maximizes the time between undesirable mode control steps. Because the parasitic mode behavior is more precisely described by backward error analysis, truncation of the smooth modified system is a more robust method of smoothing the undesired modes. It is also more easily applicable to $k-$step methods for arbitrary $k$.

The proposed re-initialization procedure does result in diffusion in phase space and in conserved quantities due to the effectively time-dependent modified system. However, whereas the parasitic mode might otherwise grow linearly or exponentially, the re-initialization procedure will result in a random walk exhibiting a root mean square distance from the original orbit proportional to $\sqrt{t}$ with a small diffusion coefficient. Mode control through re-initialization can delay the accumulation of parasitic mode errors for many numerical steps. Assuming the parasitic mode amplitude is monitored via a conserved quantity, like energy $E$, suppose the acceptable energy error threshold is $\delta_E$, where the error is relative to the most recent initialization step. The energy diffusion coefficient may be estimated as $D \approx \delta_E^2 /  \delta_t$, so it is advantageous to increase the time $\delta_t$ between re-initialization steps. By setting the threshold for re-initialization fairly small, we can also reduce $\delta_E$. Naively, one might expect the averaging step used in the Runge-Kutta smoothing/delousing to significantly reduce the effective diffusion step size $\delta_E$ compared with instantaneous re-initialization with backward error analysis. However, the oscillations are often not centered about zero, reducing the benefit of the averaging step. 

When seeking the advantages of applying a geometric integrator, such as a variational algorithm, it is desirable that any mode controlling steps retain the underlying conservation properties. Unfortunately, truncation of the smooth modified system violates preservation of the discrete symplectic structure. In the proof of Section \ref{ssec:smooth_modified_system_for_phase_space_variational_integrators}, if the flow map of the smooth modified system $\phi_\tau$ is replaced with a flow map $\phi'_\tau$ corresponding to the vector field of a truncated smooth modified system, the discrete Euler-Lagrange equations will no longer be exactly satisfied by the discrete trajectories and the errors can accumulate in the conserved symplectic structure. However, when exact preservation of the discrete symplectic structure leads to large-amplitude parasitic modes, the error introduced by the diffusive re-initialization steps is preferable to the continued amplification of parasitic modes. As will be presented in Fig.\,\ref{fig:nonlinear_pendulum}, a multistep symplectic algorithm exhibiting growing parasitic oscillations can accumulate energy error equally rapidly as a non-conservative Runge-Kutta algorithm. By actively controlling the mode via backward error re-initialization, the energy error is improved by several orders of magnitude.

\section{Numerical Examples}
\label{sec:numerical_examples}

For the first two examples, consider the explicit midpoint algorithm:
\begin{equation}
	x_{k+1} - x_{k-1} = 2 h f(x_{k}).
	\label{eq:explicit_midpoint}
\end{equation}
This algorithm is a typical test case for exhibiting parasitic modes \cite{Hairer_2006, New_1998, Vankerschaver_2013}. For canonically Hamiltonian systems, explicit midpoint applied to the equations of motion results from a discrete variational principle. To variationally derive explicit midpoint, begin with the canonical phase-space Lagrangian:
\begin{equation}
	L_{ps}(p,q, \dot{p}, \dot{q}) = p \dot{q} - H(p,q).
	\label{eq:phase_space_lagrangian}
\end{equation}
Letting $x = [p, q]$ in the above Lagrangian, a trapezoidal rule discretization of the action takes the form:
\begin{align}
	S_d(x_0, ..., x_k, ..., x_N) & \approx \int L_{ps}(x, \dot{x}) dt  = \sum_{k=0}^{N-1} L_d(x_k, x_{k+1}) \nonumber \\
	& = \sum_{k=0}^{N-1} \frac{h}{2}\left[ L_{ps}(x_k, \frac{x_{k} - x_{k-1}}{h}) + L_{ps}(x_{k+1}, \frac{x_{k+1} - x_k}{h}) \right].
\end{align}
Requiring the action to be stationary with respect to variation at arbitrary $x_k$ yields the explicit midpoint algorithm Eq.\,(\ref{eq:explicit_midpoint}).

Typically, it is ill-advised to apply a discretization of the form discussed in Ref.\,\cite{Marsden_2001} to obtain an integrator from a canonical phase space Lagrangian, precisely because it yields a two-step method for the first-order ODE and introduces parasitic oscillations. For canonical phase-space Lagrangians, alternative discretization methods exist for obtaining one-step variational algorithms \cite{Leok_2011, Lall_2006}. However, it is unknown how to obtain one-step variational integrators for first-order ODE systems resulting from non-canonical phase-space Lagrangians, so understanding the behavior of the multistep algorithms in familiar and simple canonical settings is useful for treating more advanced cases.

\subsection{Linear Oscillator}
\label{ssec:linear_oscillator}

\begin{figure}[htb]
	\center{
		\includegraphics[width=0.75\textwidth]{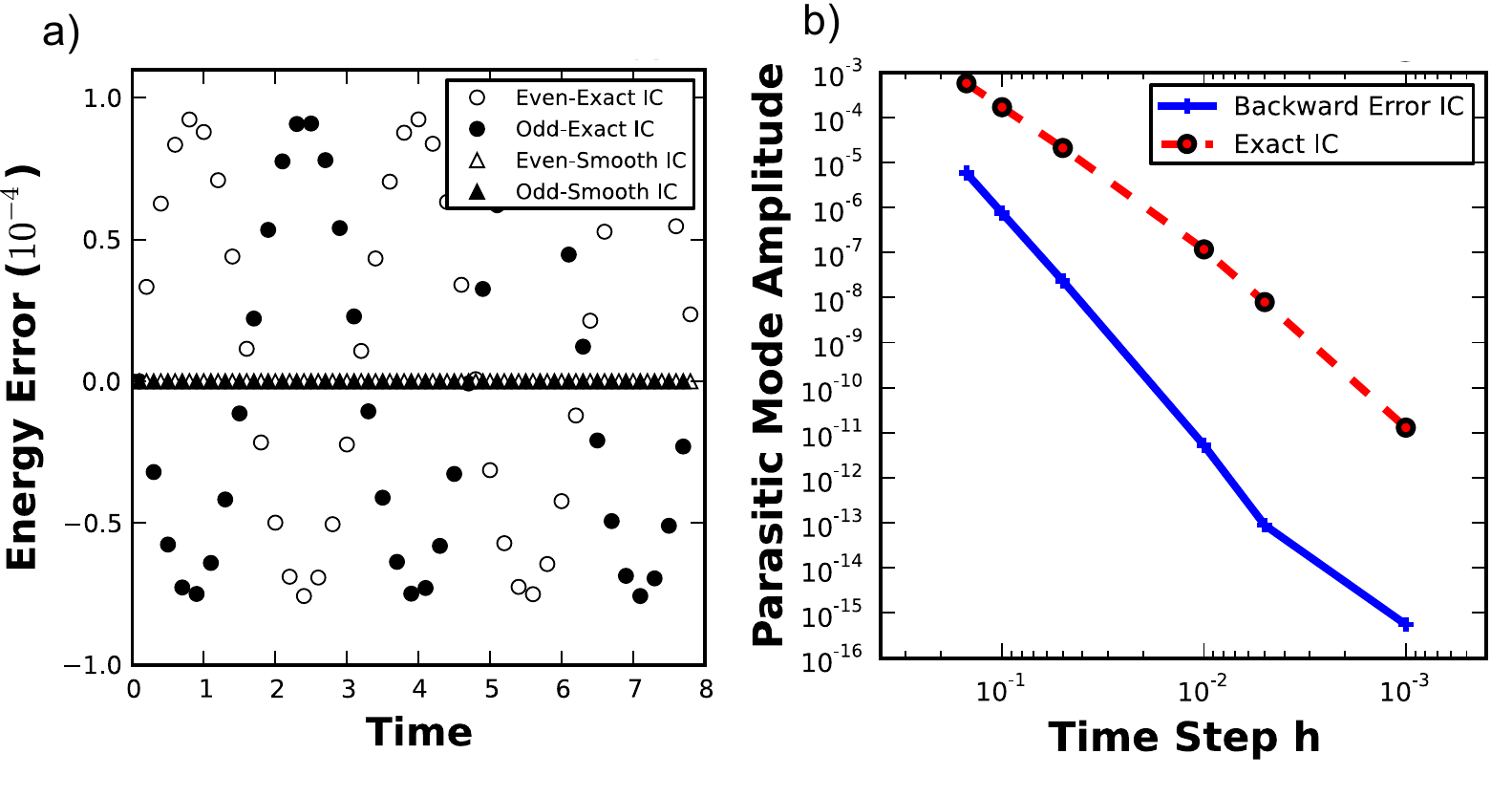}
		\caption{Comparison of initialization techniques for explicit midpoint Eq.\,(\ref{eq:explicit_midpoint}) applied to the linear oscillator Eq.\,(\ref{eq:linear_oscillator}). \textbf{a)} Parasitic even-odd mode manifest in the energy as a function of time for $x_0 = \left[  1,  0 \right]^T$, $h=0.1$. The black and white circles respectively represent even and odd steps of a single trajectory initialized with the exact solution, $x_1 = \left[  \cos(h),  -sin(h)\right]^T$. The parasitic mode is eliminated and energy is conserved energy to machine precision in the triangle trajectory which is initialized with the initial condition $x_1 = \left[  \sqrt{1-h^2} , -h \right]^T$. \textbf{b)} Comparison of parasitic mode amplitude when initial conditions are chosen by backward error analysis and the true solution. The truncated backward error initial condition is $x_1 = \left[ \cos(h + h^3/6), -\sin(h+h^3/6) \right]^T$.}
                \label{fig:linear_oscillator}
		}
\end{figure}

For instance, when explicit midpoint is applied to the linear oscillator problem:
\begin{equation}
	\left[ \begin{array}{c} \dot{x}^1 \\ \dot{x}^2 \end{array} \right] = \left[ \begin{array}{c} x^2 \\ -x^1 \end{array} \right],
	\label{eq:linear_oscillator}
\end{equation} 
the parasitic mode is oscillatory but remains bounded for all simulation time. Explicit midpoint's characteristic polynomial possesses roots at $\zeta = 1, -1$, resulting in the smooth mode and one that oscillates between positive and negative on alternating time steps due to the $\zeta_l^{t/h}$ term in Eq.\,(\ref{eq:full_modified_solution}). This behavior can be seen in Fig.\,\ref{fig:linear_oscillator} (a), which displays the energy error for explicit midpoint when initialized with the true solution to the ODE compared to explicit midpoint initialized without parasitic modes. The trajectory initialized with the non-parasitic initial conditions conserves energy to machine precision (line of triangles at 0.0), while the single parasitic mode exhibited by the black and white circles appears as two distinct trajectories. The exact ``non-parasitic" initial conditions were determined by solving for the general solution to the linear recursion relation given by the multistep method. For arbitrary nonlinear ODEs, one will not be able to calculate the exactly non-parasitic initial conditions because this would require keeping all orders of $h$ in the smooth modified system.

In Fig.\,\ref{fig:linear_oscillator} (b), the parasitic mode amplitude is compared in trajectories initialized using the true solution to the ODE and a truncated version of the smooth modified system. The numerical test confirms the $\mathcal{O}(h^5)$ scaling in the amplitude of the parasitic mode when initial conditions are chosen using backward error analysis. The parasitic mode resulting from initialization using the analytic solution only scales with $\mathcal{O}(h^3)$ due to the modified system matching the original system at only the $h^0$ and $h^1$ terms. 

\subsection{Nonlinear Pendulum}
\label{ssec:nonlinear_pendulum}

\begin{figure}[htb]
	\center{
		\includegraphics[width=0.75\textwidth]{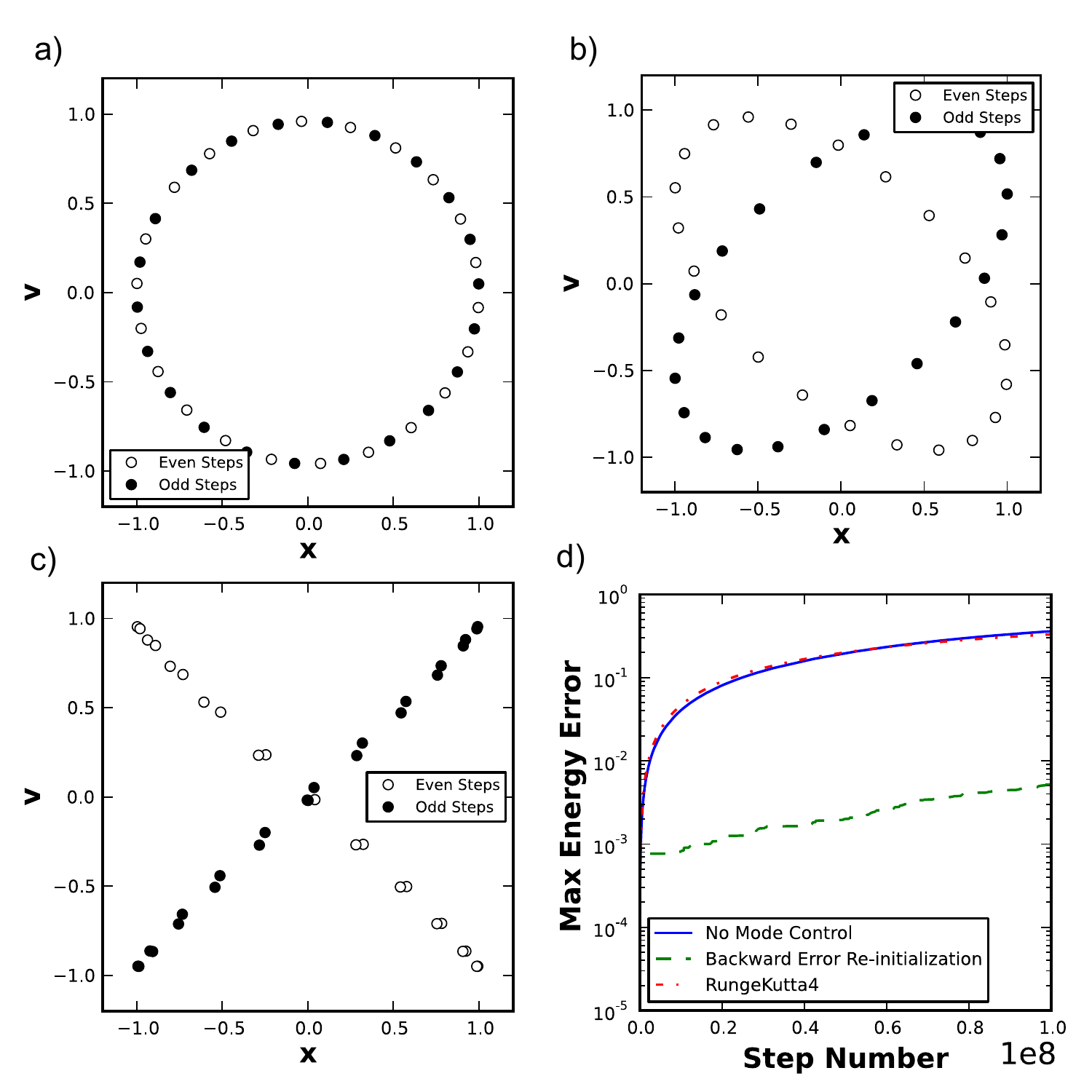}
		\caption{ Explicit midpoint applied to the nonlinear pendulum problem, Eq.\,(\ref{eq:nonlinear_pendulum}), with step size $h=0.15$ and initialized using fourth-order Runge-Kutta for parts \textbf{a, b, c}. \textbf{a)} At early times, such as after $1\times10^4$ steps, the phase portrait is reasonable and the parasitic mode is only observable at very small scales. \textbf{b)} After $8\times10^5$, the growing parasitic mode greatly distorts the phase-space trajectory. The even- and odd-numbered time steps have diverged onto distinct trajectories.  The accumulated energy error is 0.25. \textbf{c)} After $2\times10^6$ steps, the energy error reaches a maximum amplitude of 0.46 and the phase portrait has completely deteriorated. \textbf{d)} Demonstration of controlling the parasitic mode by re-initializing the system when the energy error reaches $5\times10^{-4}$ compared to the most recent initialized energy. The maximum energy error on the $y-$axis refers to the maximum energy error in all preceding steps. The step size was $h=0.1$.}
                \label{fig:nonlinear_pendulum}
		}
\end{figure}

Next, we apply explicit midpoint to the nonlinear pendulum problem: 
\begin{equation}
	\left[ \begin{array}{c} \dot{x}^1 \\ \dot{x}^2 \end{array} \right] = \left[ \begin{array}{c} x^2 \\ -\sin(x^1) \end{array} \right].
	\label{eq:nonlinear_pendulum}
\end{equation} 
The nonlinearity introduces more complicated behavior in the parasitic mode, as shown in Fig.\,\ref{fig:nonlinear_pendulum}. While the parasitic root is linearly marginally stable, the amplitude is observed to grow slowly in time. Analysis of the instability requires solving the full modified system Eq.\,(\ref{eq:full_modified_system}).  At short times the mode grows linearly, observable only at small scales in the phase-space trajectory (Fig.\,\ref{fig:nonlinear_pendulum} (a)) or in the energy error. At moderate times, clearly distinct trajectories emerge for the even- and odd-step solutions (Fig.\,\ref{fig:nonlinear_pendulum} (b)), and the energy error is dominated by the parasitic mode. At maximum energy error, the phase plot no longer resembles the true solution, but instead diagonal lines formed by the even and odd trajectories as can be seen in Fig.\,\ref{fig:nonlinear_pendulum}(c). Eventually, the mode oscillates with a slow period and large amplitude, restoring a more reasonable phase-space trajectory. The maximum amplitude of the mode is observed to be the same regardless of initial conditions, however, the early-time linear growth rate and the period of the slow oscillation depend on the initial amplitude of the parasitic mode and therefore on the full set of initial conditions. At $h=0.15$, the backward error initialized mode had a period 100 times longer than the true solution initialization.

Numerical results using both backward error initialization and mode stabilization are given in Fig.\,\ref{fig:nonlinear_pendulum}(d). Without re-initialization, the energy error accumulated via the parasitic mode grows at nearly the same rate as energy error accumulates for a fourth-order Runge-Kutta integrator. However, this can be partially mitigated by re-initializing whenever the energy error becomes noticeably dominated by the parasitic mode. This results in a random walk in energy, the maximum deviation of which is plotted in the bottom right hand figure of Fig. \ref{fig:nonlinear_pendulum}. By initializing the parasitic mode to small amplitude using backward error initialization and by periodically re-initializing the system, diffusion in energy takes extremely long to accumulate and explicit midpoint yields a reasonable numerical solution for very long simulation times despite the parasitic instability.

\subsection{Planar Point Vortices}
\label{ssec:planar_point_vortices}

Finally, we apply the backward error initialization procedure to systems of planar point vortices. For $n$ point vortices interacting in the $x-y$ plane, the equations of motion are given by \cite{Aref_2007}:
\begin{align}
	\dot{x_i} & = -\frac{1}{2\pi} \sum_{j=1, \ne i}^{N} \frac{\Gamma_j (y_i - y_j)}{ (x_i - x_j)^2 + (y_i - y_j)^2}, \nonumber \\ 
	\dot{y_i} & = \frac{1}{2\pi} \sum_{j=1, \ne i}^{N} \frac{\Gamma_j (x_i - x_j)}{ (x_i - x_j)^2 + (y_i - y_j)^2}, \quad \quad i = 1,2,...,n
	\label{eq:eom_point_vortices}
\end{align}
where the $\Gamma_j$ are constants and the summation does not include the singular term arising when $i=j$. To demonstrate backward error initialization for a not linear multistep algorithm, we apply a ``two-step midpoint'' algorithm. Letting $z = [x, y]$, the method is given by: 
\begin{equation}
	z_{k+1} - z_{k-1} = h\left[f(\frac{z_{k-1} + z_{k}}{2}) + f(\frac{z_{k} + z_{k+1}}{2})\right].
	\label{eq:two-step_midpoint}
\end{equation}
This algorithm is again variational, corresponding to a midpoint discretization of the phase-space Lagrangian Ref.\,\cite{Rowley_2002}:
\begin{equation}
  L_d(z_k, z_{k+1}) = h L_{ps}(\frac{z_k + z_{k+1}}{2}, \frac{z_{k+1} - z_k}{h}).
\end{equation}
This algorithm is simply the implicit midpoint scheme composed with itself, and exhibits stable parasitic modes when applied to the planar point vortices problem \cite{Vankerschaver_2013}.

While two-step midpoint is not a linear multistep algorithm but instead a more general symmetric multistep method, backward error initialization improves the smoothness of the conserved quantities by reducing the amplitude of the parasitic modes. Rigorous estimates of long-time behavior of these more general multistep methods have recently been presented in Ref.\,\cite{DAmbrosio_2013}. We calculate the  $\mathcal{O}(h^2)$ smooth modified system for this algorithm to be:
\begin{equation}
	\dot{\tilde{x}} = f(\tilde{x}) + h^2 \left[\frac{-1}{24} f''(\tilde{x}) f(\tilde{x}) f(\tilde{x}) + \frac{1}{12} f'(\tilde{x}) f'(\tilde{x}) f(\tilde{x}) \right].
	\label{eq:two-step_midpoint_f3}
\end{equation}

\begin{figure}[htb]
	\center{
		\includegraphics[width=0.75\textwidth]{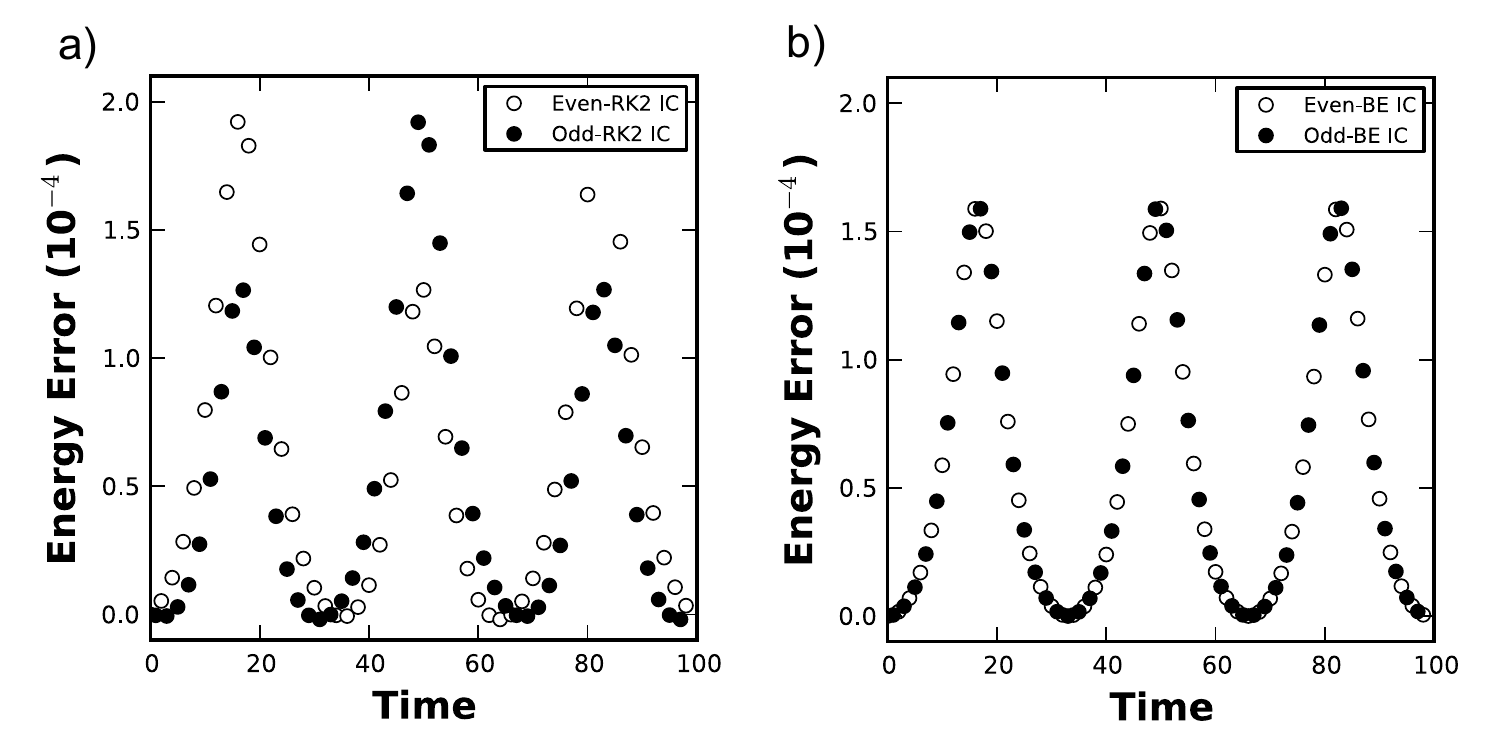}
		\caption{ Two-step midpoint Eq.\,(\ref{eq:two-step_midpoint}) applied to a system of four planar point vortices described by Eq.\,(\ref{eq:eom_point_vortices}). Both figures used a first initial condition of $z_0 = [x^1_0, x^2_0, ..., y^1_0, y^2_0, ...] = [-1, 1, -1, 1, 2, 2, -2, -2]$ and a step size of $h=1$. \textbf{a)} Two-step midpoint trajectory initialized using second-order Runge-Kutta as in Ref.\,\cite{Rowley_2002}. \textbf{b)} Two-step midpoint trajectory initialized using the truncated smooth modified system of Eq.\,(\ref{eq:two-step_midpoint_f3}) }
                \label{fig:point_vortices}
		}
\end{figure}

Figure \ref{fig:point_vortices} compares two-step midpoint trajectories initialized using second-order Runge-Kutta (as in Ref.\,\cite{Rowley_2002}) with backward error initialized trajectories. Better approximations of the true solution exhibit less severe parasitic oscillations, but the parasitic modes remain apparent on this scale of energy error even for excellent approximations of the true solution. In contrast, the numerical trajectory of the smooth modified system possesses an undetectably small parasitic oscillation. The parasitic mode amplitudes are identified by comparing the test trajectory to a trajectory initialized using the underlying one-step method, implicit midpoint. In general, however, one will not know the underlying one-step method, so backward error initialization provides a practical and generally applicable tool for prescribing initial conditions that yield smooth trajectories.

\section{Conclusion}
\label{sec:conclusion}

Backward error analysis yields insightful explanations of the behavior of multistep methods. Aside from understanding the stability and presence of parasitic modes, it also guides the selection of superior sets initial conditions for achieving the desired numerical trajectory. The standard for multistep methods is initialization using approximations of the true solution generated by one-step methods. It is our belief that the advantages that can be gained by approximating the truncated smooth modified system are worth the marginal effort required to obtain them. Algorithms possessing a variational formulation on a phase-space Lagrangian, in particular, will behave better when initialized using backward error analysis. Control of the parasitic modes is critical for taking advantage of the good long term behavior of variational integrators applied to non-canonical Hamiltonian systems, and smooth initial conditions combined with re-initialization steps are powerful techniques for enhancing the long term numerical fidelity.

Of course, the method is not without limitations and disadvantages. For one, initialization using backward error analysis requires implementation of higher-order derivatives of the ODE function than would otherwise be necessary. This drawback increases with increasingly accurate multistep systems. For a method accurate to $\mathcal{O}(h^p)$, the first non-zero function in the modified equation will be $f_{p+1}$, and will typically depend on all derivatives $f^{(n)}$ with $n \leq p$. Symbolic or automatic differentiation tools are almost certainly necessary for this task unless $n$ is small, and might require the inclusion of otherwise unneeded packages or libraries in the ODE solving program. Another drawback is the potentially slow speed of the initialization and mode control re-initialization steps. Accurate solution of the modified system using a one-step method may require sub-stepping or a Runge-Kutta method with many function evaluations compared to the single function evaluation used in the multistep method. Given the rarity of initialization or re-initialization compared to the typical multistep update, this should not significantly impact total computation time.

These drawbacks aside, the modeling of dynamical systems using multistep methods can likely be improved in a variety of application domains by considering the smooth modified system derived using backward error analysis. 

This work was supported by DOE contract number DE-AC02-09CH11466 and a Fusion Energy Sciences Fellowship. The authors are grateful to Jonathan Squire, Nawaf Bou-Rabee, and Samuel Lazerson for helpful discussions on this topic. 

\bibliographystyle{elsarticle-num}
\bibliography{SI_refs}

\end{document}